\documentclass{iopart}
\usepackage{iopams}

\usepackage[utf8]{inputenc}
\usepackage{bm}
\usepackage{xspace}

\usepackage{ifthen}

\newcommand{\quotes}[1]{``#1''}
\newcommand{\eqref}[1]{(\ref{#1})}

\newcommand{\myequation}{\begin{equation}}
\newcommand{\myendequation}{\end{equation}}
\let\[\myequation
\let\]\myendequation

\newcommand{\ZZ}{\cc{Z}}
\newcommand{\Deriv}[2][\empty]{%
  \ifthenelse{\equal{#1}{\empty}}
    {D_#2}
    {D_{#1,#2}}
}
\newcommand{\DDeriv}[1]{\vec{D}_{#1}}
\newcommand{\psit}[1][\empty]{%
  \ifthenelse{\equal{#1}{\empty}}
    {\psi_t}
    {\psi_t^{(#1)}}
}
\newcommand{\npsit}[1][\empty]{%
  \ifthenelse{\equal{#1}{\empty}}
    {\tilde\psi_t}
    {\tilde\psi_t^{(#1)}}
}
\newcommand{\rhot}[1][\empty]{%
  \ifthenelse{\equal{#1}{\empty}}
    {\rho_t}
    {\rho_t^{(#1)}}
}

\renewcommand{\vec}[1]{\boldsymbol{#1}}
\newcommand{\bra}[1]{\langle #1 |}
\newcommand{\ket}[1]{| #1 \rangle}
\newcommand{\braket}[2]{\langle #1 | #2 \rangle}

\newcommand{\ii}{\mathrm{i}}
\newcommand{\dd}{\, \mathrm{d}}

\newcommand{\ee}{\vec{e}}
\newcommand{\kk}{\vec{k}}
\newcommand{\mm}{\vec{m}}
\newcommand{\nn}{\vec{n}}
\newcommand{\ww}{\vec{w}}
\newcommand{\zz}{\vec{z}}
\renewcommand{\Tr}{\mathrm{Tr}}
\renewcommand{\H}[1]{H_\mathrm{#1}}

\newcommand{\cc}[1]{{#1}^*}
\newcommand{\abs}[1]{{\left\vert #1 \right\vert}}
\newcommand{\adj}[1]{{#1}^\dagger}
\newcommand{\E}[1][\empty]{%
  \ifthenelse{\equal{#1}{\empty}}
    {\mathbb{E}}
    {\mathbb{E}\left( #1 \right)}
}
\renewcommand{\exp}[1][\empty]{%
  \ifthenelse{\equal{#1}{\empty}}
    {\mathrm{exp}}
    {\mathrm{e}^{#1}}
}

\newcommand{\Texp}[1][\empty]{%
  \ifthenelse{\equal{#1}{\empty}}
    {\operatorname{T}_+\mathrm{exp}}
    {\operatorname{T}_+\mathrm{e}^{#1}}
}
\newcommand{\Comm}[1]{\left[#1\right]}
\usepackage{ulem}  
\usepackage{color}

\begin{document}
\title[Hierarchical equations for open system dynamics]{Hierarchical equations for open system dynamics in fermionic and bosonic environments}

\author{D.\ Suess$^{1,3}$, W.\ T.\ Strunz$^2$, and A.\ Eisfeld$^1$}

\address{${}^1$ Max Planck Institute for Physics of Complex Systems, N\"othnitzer Stra\ss e 38, D-01187 Dresden, Germany}
\address{${}^2$ Institut f\"ur Theoretische Physik, Technische Universit\"at Dresden, D-01062 Dresden, Germany}
\address{${}^3$ Institute of Physics, University of Freiburg, Rheinstra\ss e 10, D-79104 Freiburg, Germany}
\ead{daniel.suess@physik.uni-freiburg.de}

\date{\today}
\begin{abstract}
  We present novel approaches to the dynamics of an open quantum system coupled linearly to a non-Markovian fermionic or bosonic environment.
  In the first approach, we obtain a hierarchy of stochastic evolution equations of the diffusion type.
  For the bosonic case such a hierarchy has been derived and proven suitable for efficient numerical simulations recently [arXiv:1402.4647].
  The stochastic fermionic hierarchy derived here contains Grassmannian noise, which makes it difficult to simulate numerically due to its anti-commutative multiplication.
  Therefore, in our second approach we eliminate the noise by deriving a related hierarchy of density matrices.
  A similar reformulation of the bosonic hierarchy of pure states to a master equation hierarchy is also presented.
\end{abstract}

\section{Introduction}
\label{sec:Intro}

The theory of open quantum systems has become an important topic in modern physics and its applications, since many interesting phenomena of quantum systems emerge only when the influence of the system's environment is taken into account~\cite{weiss_quantum_2012,MaKue00__,breuer_theory_2007}.
However, realistic system-bath models are often analytically and numerically intractable even if one is only interested in the relevant degrees of freedom, i.e.\ the reduced density operator.
Nevertheless, since the low-energy behavior of the environment is generally independent of its dynamical details~\cite{wilson_renormalization_1975}, a coarse-grained description is often sufficient for all practical purposes.
One standard example is an environment of uncoupled harmonic oscillators, which is a good approximation for weakly coupled delocalized modes~\cite{feynman_theory_1963}.
Here, we will address another standard model, namely an environment of non-interacting fermions.

These environmental models are often simplified further using the Born-Markov approximation for a \quotes{memory-less}, weakly coupled environment~\cite{breuer_theory_2007}.
Recently, there has been a growing interest in going beyond this Markov approximation, which fails in particular if the coupling between system and bath is not weak or when one deals with a structured environment~\cite{weiss_quantum_2012,MaKue00__}.
Because of its importance, e.g., for transport through molecules or quantum dots, various approaches have been developed to go beyond the Born-Markov regime in fermionic environments (see for example \cite{JiZhYa08_234703_,Ti08_195416_,CrSa09_245311_,PoKl13_2288_}).
Distinguished features of non-Markovian dynamics are discussed in~\cite{zhang_general_2012}.

One powerful approach to obtain the dynamics of the system for such a non-Markovian setting is non-Markovian quantum state diffusion (NMQSD).
The NMQSD approach was originally derived for a harmonic oscillator environment~\cite{strunz_linear_1996,Di96_309_,diosi_non-markovian_1997,diosi_non-markovian_1998}.
A related NMQSD equation for fermionic environments has been derived recently~\cite{zhao_fermionic_2012,chen_non-markovian_2013}.
Within NMQSD one obtains a stochastic Schrödinger equation of the diffusion type that lives in the Hilbert space of the system.
Its solutions---so called quantum trajectories---yield the reduced density operator by an average over distinct realizations.
The main difference between the bosonic and fermionic theory is the noise processes entering the equations of motion:
While the influence of a bosonic environment can be described exactly by a complex-valued colored Gaussian process, fermionic environments require the use of Grassmannian colored noise.

An obstacle of the NMQSD equations is that it contains the noise not only as a multiplicative term, but also within a functional derivative under a memory integral.
To tackle this problem, we have recently derived a hierarchy of pure states (HOPS) for the bosonic case~\cite{sus_hierarchy_2014}.
This hierarchy consists of a set of coupled equations where the noise enters only linear.
The price one has to pay is that instead of an intractable memory integral one deals with an infinite hierarchy of coupled stochastic equations.
Fortunately, it turns out that one can often truncate the hierarchy at quite low order resulting in a system of equations which can be solved numerically in an efficient way.

One main result of the present work is the derivation of such a hierarchy of pure states for the fermionic theory, which has structure very similar to the bosonic one.
In contrast to the bosonic case, where one can easily generate the complex Gaussian noise, a numerical simulation of the fermionic HOPS seems to be unfeasible due to the anti-commuting Grassmannian processes.
Therefore, we will go one step further and eliminate the noise by deriving a hierarchy of master equations based on the fermionic HOPS, which can be solved numerically efficient.
The corresponding master equation for a bosonic environment based on the established hierarchy of pure states~\cite{sus_hierarchy_2014} is also presented.
Although the hierarchical description of pure state dynamics is quite new, hierarchical equations of motion for density operators of open quantum systems coupled to bosonic~\cite{tanimura_stochastic_2006} or fermionic~\cite{JiZhYa08_234703_} environments are well established tools.\\

The paper is organized as follows:
In Sec.~\ref{sec:fermionic} we elaborate on the fermionic theory:
First, we recapitulate the fermionic NMQSD approach in Sec.~\ref{sub:fermionic_nmqsd} in order to recall the established theory and introduce our notation.
Then, Sec.~\ref{sub:fermionic_hierarchy} is devoted to the fermionic HOPS and the hierarchy of density matrices is derived in the following Sec.~\ref{sub:master_equation_hierarchy}.
Finally, we discuss finite-temperature environments in~\ref{sub:thermal}.
The corresponding theory for bosonic environments is treated in Sec.~\ref{sec:bosonic_environments}:
Sec.~\ref{sub:bosonic_hierarchy_of_pure_states} summarizes the HOPS construction from~\cite{sus_hierarchy_2014}.
The novel result for bosonic environments, namely the hierarchy of master equations, can be found in Sec.~\ref{sub:bosonic_master_equation_hierarchy}.
We use units where $k_{\rm B}=\hbar=1$.

\section{Fermionic Environments}
\label{sec:fermionic}

Let us consider the (total) Hamiltonian
\[
  \H{tot} = H + \H{env} + \H{int},
  \label{eq:h_tot}
\]
where $H$ captures the system's free dynamics and
\(
  \H{env} = \sum_{j, \lambda} \omega_{j,\lambda} \adj{b}_{j,\lambda} b_{j,\lambda}
\)
the dynamics of an environment consisting of indistinguishable spin-1/2 particles described by fermionic ladder operators obeying canonical anti-commutator relations
\[
  \Big\{ b_{j,\lambda}, b_{j',\lambda'} \Big\} = 0
  \quad\mbox{and}\quad
  \Big\{ b_{j,\lambda}, \adj{b}_{j',\lambda'} \Big\} = \delta_{jj'} \delta_{\lambda\lambda'}.
  \label{eq:anticommutators}
\]
The interaction of system and environment is modeled by a linear coupling Hamiltonian
\[
  \H{int} = \sum_{j,\lambda} \left( \cc{g}_{j,\lambda} L_j \adj{b}_{j,\lambda} + g_{j,\lambda} \adj{L}_j b_{j,\lambda} \right).
\]
Here, $L_j$ are system operators and $g_{j,\lambda}$ are complex numbers quantifying the coupling strength of the respective fermion $(j,\lambda)$.
We assume that all system operators commute with environment operators or, put differently, that the system is distinguishable from the environment.
Such a model arises, for example, in the description of tunneling through a quantum dot or molecules~\cite{Ti08_195416_}.
It is convenient to encode the frequency dependence of the interaction strengths in the so called spectral densities
\[
  J_j(\omega) = \sum_{\lambda} |g_{j,\lambda}|^2 \delta(\omega-\omega_{j,\lambda}),
  \label{eq:spectral_density}
\]
which are typically assumed to be continuous functions of frequency.\\

For now we will confine the discussion to the zero-temperature case with pure initial condition
\[
  \label{eq:initial_Psi}
  \ket{\Psi_0} = \ket{\psi_0}\otimes\ket{0}
\]
and treat the more general case of a thermal initial state in Sec.~\ref{sub:thermal}.
Here, $\ket{0}$ denotes the vacuum with respect to all $b_{j,\lambda}$.
Since the full dynamics governed by the Hamiltonian~\eqref{eq:h_tot} is unitary, the full state of system and bath can be described by a pure state $\ket{\Psi_t}$ at all times.
The reason for introducing a distinction between system and environment in the first place is that we are only interested in the reduced state of the former
\[
  \rho(t)=\Tr_\mathrm{env}{\ket{\Psi_t}\bra{\Psi_t}}.
  \label{eq:partial_trace}
\]
However, the NMQSD formalism recalled in the next section is formally equivalent to solving the full Schrödinger equation for $\ket{\Psi_t}$.

\subsection{Fermionic NMQSD}
\label{sub:fermionic_nmqsd}

The theory of non-Markovian quantum state diffusion for fermionic environments has been derived in~\cite{zhao_fermionic_2012,chen_non-markovian_2013}.
Here, we will briefly recapitulate the crucial steps in order to establish the notation used throughout the paper.

Similar to the bosonic case, the fermionic NMQSD theory is based on a representation of the bath degrees of freedom in coherent states
\(
  \ket{\zz} := \mathop{\otimes}_{j}\mathop{\otimes}_{\lambda}\ket{z_{j\lambda}},
\)
where a (non-normalized) fermionic coherent state is defined similarly to its bosonic counterpart by
\[
  \ket{z_{j\lambda}} =\exp[-z_{j\lambda} \adj{b}_{j\lambda}] \ket{0} = \ket{0} - z_{j\lambda} \adj{b}_{j,\lambda} \ket{0}
\]
Here, the $z_{j\lambda}$ are anti-commuting Grassmann variables with
\(
  \{z_{j\lambda},z_{j'\lambda'}\}=\{\cc{z}_{j\lambda},z_{j'\lambda'}\}=\delta_{j,j'}
\)
and
\(
  \{z_{j\lambda},b_{j'\lambda'}\}=\{z_{j\lambda},b^{\dagger}_{j'\lambda'}\}=\delta_{j,j'}.
\)
For more details on these coherent states see e.g.~\cite{cahill_density_1999,combescure_fermionic_2012}.

We now expand the bath degrees of freedom of the full system-environment state with respect to the coherent states introduced above
\(
  \psit(\cc\zz) := \braket{\zz}{\Psi_t}.
\)
We also absorb the free time evolution of the environment using the interaction picture with respect to $\H{env}$.
The resulting Schrödinger equation for $\psit(\cc\zz)$ then reads
\begin{eqnarray}
  \partial_t \psit(\cc\zz)
  &=& -\ii H \psit(\cc\zz)
  -\ii \sum_{j,\lambda} \cc{g}_{j,\lambda} L_j \exp[\ii\omega_{j,\lambda}t] \cc{z}_{j,\lambda} \psit(\cc\zz) \nonumber\\
  &&-\ii \sum_{j,\lambda} g_{j,\lambda} \adj{L}_j \exp[-\ii\omega_{j,\lambda}t] \overrightarrow{\partial}_{\cc{z}_{j,\lambda}} \psit(\cc\zz),
  \label{eq:schroedinger_ia}
\end{eqnarray}
where  $\overrightarrow{\partial}_{\cc{z}_{j,\lambda}}$ denotes the left-derivative with respect to $\cc{z}_{j,\lambda}$.
In the following we will drop the arrow from left-derivatives (and only indicate right-derivatives explicitly).

The main result of~\cite{zhao_fermionic_2012,chen_non-markovian_2013} is that~\eqref{eq:schroedinger_ia} can be recast into a \emph{stochastic Schrödinger equation} of the quantum state diffusion type, namely
\begin{eqnarray}
  \partial_t \psit(\ZZ)
  &=& -\ii H \psit(\ZZ)
  + \sum_j L_j \ZZ_j(t) \psit(\ZZ) \nonumber\\
  &&- \sum_j \adj{L}_j \int_0^t \alpha_j(t-s) \frac{\delta \psit(\ZZ)}{\delta\ZZ_j(s)} \dd s
  \label{eq:nmsse}
\end{eqnarray}
where the stochastic process
\(
  \ZZ_j(t) = -\ii \sum_\lambda \cc{g}_{j,\lambda} \exp[\ii\omega_{j,\lambda} t] \cc{z}_{j,\lambda}
\)
is characterized by its correlation function
\[
  \alpha_j(t) = \sum_\lambda \abs{g_{j,\lambda}}^2 \exp[-\ii\omega_{j,\lambda} t]
  \label{eq:vacuum_bcf}
\]
through
\[
  \E Z_j(t) = \E[Z_j(t)Z_{j'}(s)] = 0,
  \quad
  \E[Z_j(t) \ZZ_{j'}(s)] = \delta_{jj'}\alpha(t-s).
  \label{eq:correlations}
\]
Note that the $\ZZ_t$ are \emph{Grassmannian processes}, i.e.~values at different times anti-commute hindering an efficient numerical generation of these processes.
Furthermore, Eq.~\eqref{eq:vacuum_bcf} is the well-known relation of the zero-temperature bath correlation function and the spectral density~\eqref{eq:spectral_density} through a (one-sided) Fourier transform~\cite{weiss_quantum_2012}.

Besides describing the correlation of the noise process, the bath correlation function also weights the functional derivative at different times under the memory integral in~\eqref{eq:nmsse}.
Since this term is non-local in time as well as in the realization of the processes, it is unclear how this term can be evaluated in general.
We will abbreviate the full memory integral introducing the (left-)derivation operator\footnote{%
  Note that the integral boundaries in~\eqref{eq:nmsse} arise from the specific vacuum initial conditions~\eqref{eq:initial_Psi}; see \cite[Footnote 42]{sus_hierarchy_2014} for details.
  \label{fn:integral_boundaries}
}
\[
  \Deriv[j]{t} \psit(\ZZ) = \int \alpha_j(t-s) \frac{\delta\psit(\ZZ)}{\delta \ZZ_j(s)} \dd s.
  \label{eq:derivation_operator}
\]
In previous works~\cite{zhao_fermionic_2012,yu_non-markovian_2004}, this functional derivative was replaced by an operator ansatz acting in the system's Hilbert space
\(
  \delta \psit(\ZZ) / \delta \ZZ_j(s) = Q(t,s,\ZZ) \psit(\ZZ).
\)
For certain simple models, this ansatz is exact and the $Q$-operator independent of the noise~\cite{zhao_fermionic_2012}.
The latter property can be used to derive a master equation for the reduced density operator~\eqref{eq:partial_trace}.
However, no feasible scheme for calculating the $Q$-operator and, hence, solving the fermionic NMQSD equation~\eqref{eq:nmsse} has been found so far.
Therefore, we will present a different approach in the next section that does not rely on the $Q$-operator ansatz.

\subsection{Fermionic Hierarchy of Pure States}
\label{sub:fermionic_hierarchy}

In analogy to the bosonic case of Ref.~\cite{sus_hierarchy_2014} we define auxiliary states using the derivation operators~\eqref{eq:derivation_operator}
\[
  \psit[\kk] := \Deriv[1]{t}^{k_1} \Deriv[2]{t}^{k_2} \ldots \psit = \DDeriv{t}^{\kk} \psit.
  \label{eq:aux_states}
\]
In contrast to HOPS for a harmonic oscillator environment, the order of functional derivative operators is relevant here, since they anti-commute.
For the same reason, all auxiliary states with some $k_j > 1$ vanish as expected from the fact that the $\Deriv[i]{t}$ are linear combinations of fermionic annihilation operators from the microscopic point of view~\eqref{eq:schroedinger_ia}.
With this notation the memory integrals in~\eqref{eq:nmsse} can be written as
\[
  \psit[1,0,\ldots] := \Deriv[1]{t} \psit, \quad
  \psit[0,1,\ldots] := \Deriv[2]{t} \psit, \quad
  \ldots
  \label{eq:first_orders}
\]
Identifying $\psit$ with $\psit[\boldsymbol{0}]$  allows us to rewrite~\eqref{eq:nmsse} as
\[
  \partial_t \psit[0] = -\ii H \psit[0] + \sum_j L_j \ZZ_j(t) \psit[0] - \sum_j \adj{L}_j \psit[\ee_j].
  \label{eq:hierarchy_zeroth_order}
\]
Here, we introduce a more compact notation of~\eqref{eq:first_orders} using the $j$-th unit vector in $\mathbb{R}^J$ with $J$ being the number of processes in~\eqref{eq:nmsse}.

In order to obtain the equations of motion for the auxiliary states~\eqref{eq:aux_states} we assume that the  bath correlation functions have an exponential form
\[
  \alpha_j(t) = g_j \exp[-\gamma_j \abs{t} - \ii\Omega_j t].
  \label{eq:exponential_bcf}
\]
In the following we use the short hand $w_j = \gamma_j + \ii\Omega_j$ for the exponent and refer to the tuple $(g_j,w_j)$ as a \quotes{mode}.
It is easy to generalize to a sum of exponentials similar to the bosonic case~\cite{sus_hierarchy_2014}.
Such bath correlation functions arise naturally for many models in the finite-temperature case $T>0$ as discussed in Sec.~\ref{sub:thermal}.

For such exponential BCFs we obtain (see~\ref{sec:derivation_of_hierarchy} for the details) the hierarchy
\begin{eqnarray}
  \partial_t \psit[\kk]
  &=& \left( -\ii H - \kk\cdot\ww + (-1)^\abs{\kk} \, \sum_j \ZZ_j(t) L_j \ \right) \psit[\kk] \nonumber\\
  &&+ \sum_j (-1)^{\abs{\kk}_j} g_j L_j \psit[\kk - \ee_j]
  - \sum_j (-1)^{\abs{\kk}_j} \adj{L}_j \psit[\kk + \ee_j]
  \label{eq:pure_hierarchy}
\end{eqnarray}
Here, we have introduced $\kk=(k_1,\dots,k_J)$, $\ww=(w_1,\dots,w_J)$, and $\kk\cdot\ww = k_1 w_1 + \cdots + k_J w_J$.
Furthermore, we use the notation $\abs{\kk} = k_1 + \cdots + k_J$ for the sum over all $k_j$ and $\abs{\kk}_j = k_{j+1} + \cdots + k_J$ denotes the sum \emph{without} the first $j$ components.
In~\eqref{eq:pure_hierarchy} all states with some $k_j \notin \{0,1\}$ vanish as mentioned below Eq.~\eqref{eq:aux_states}.
The initial conditions \eqref{eq:initial_Psi} translate to
\[
  \psi^{(0)}_0 = \psi_0 \quad \mbox{ and } \quad \psi^{(\kk)}_0 = 0 \mbox{  for  } \kk \neq 0.
\]
Equation \eqref{eq:pure_hierarchy} is our first important result of this paper.
It will serve as a starting point for the derivation of a density matrix hierarchy.

Note that~\eqref{eq:pure_hierarchy} is a \emph{finite} system of $2^{J}$ coupled equations (as before, $J$ denotes the number of modes).
This is in marked contrast to the bosonic case of Ref.~\cite{sus_hierarchy_2014}, where one has an \emph{infinite} system of equations, which has to be truncated for all practical purposes.
Although now the system \eqref{eq:pure_hierarchy} is finite, for a large number of modes $J$ it quickly becomes intractable large for numerical simulations.
Therefore, as in the bosonic case, it is useful to truncate the system in an appropriate way.
One possibility would be to exclude all states with $ |\kk| > \mathcal{K}$, where $\mathcal{K}$ is called the truncating order.
Comparing calculations with increasing order allows for a systematic check of convergence.
Another possible truncation criterion is $ |\ww \cdot \kk| > \mathcal{W}$, where $\mathcal{W}$ is a \quotes{maximal} energy.
However, even with a suitable truncation, direct simulation as for the bosonic HOPS is problematic since Grassmannian stochastic processes are hard to simulate.

\subsection{Fermionic Master Equation Hierarchy}
\label{sub:master_equation_hierarchy}

From the stochastic trajectories we obtain the reduced density operator by~(see~Ref.~\cite{zhao_fermionic_2012}, Eq.~(17))
\[
  \rhot = \E[ \ket{\psit(\ZZ)}\bra{\psit(-\ZZ)} ]= \E[ \ket{\psit(\ZZ)}\bra{\npsit(\ZZ)} ],
  \label{eq:reduced_density}
\]
where we have defined
\(
  \npsit(\ZZ) = \psit(-\ZZ).
\)
Put into words: $\npsit(\ZZ)$ is the solution of~\eqref{eq:nmsse} evaluated at $-\ZZ$.
Therefore, for each realization $\psit(\ZZ)$ one would have to compute $\psit(-\ZZ)$ as well by propagating~\eqref{eq:pure_hierarchy} with $\ZZ$ replaced by $-\ZZ$.

Our aim is to get rid of the problematic noise processes in~\eqref{eq:pure_hierarchy} by constructing a hierarchy of density operators similar to that for the pure states \eqref{eq:aux_states}.
To this end, we introduce auxiliary density operators by
\[
  \rhot[\mm,\nn] = \E[ \ket{\psit[\mm]}\bra{\npsit[\nn]} ]
  \label{eq:def_rho}
\]
with $\npsit[\nn](\ZZ) = \psit[\nn](-\ZZ)$.
Using the equations of motion~\eqref{eq:pure_hierarchy} we find
\begin{eqnarray}
    \partial_t \rhot[\mm,\nn]
    &=& -\ii \Comm{H, \rhot[\mm,\nn]} - (\mm\cdot\ww + \nn\cdot\cc{\ww}) \rhot[\mm,\nn] \nonumber\\
    &&+ (-1)^\abs{\mm} \sum_j L_j \E[ \ZZ_j(t) \ket{\psit[\mm]}\bra{\npsit[\nn]} ] \nonumber\\
    &&+ (-1)^\abs{\nn} \sum_j \E[ \ket{\psit[\mm]}\bra{\npsit[\nn]} (-Z_j(t))] \adj{L}_j \nonumber\\
    &&+ \sum_j \left( (-1)^{\abs{\mm}_j} g_j L_j \rhot[\mm-\ee_j,\nn] + (-1)^{\abs{\nn}_j} \cc{g}_j \rhot[\mm,\nn-\ee_j] \adj{L}_j \right) \nonumber\\
    &&- \sum_j \left( (-1)^{\abs{\mm}_j} \adj{L}_j \rhot[\mm+\ee_j,\nn] + (-1)^{\abs{\nn}_j} \rhot[\mm,\nn+\ee_j] L_j \right)
  \label{eq:master_hierarchy_almost}
\end{eqnarray}
This equation still contains the Grassmann processes in the averages.
However, these can be eliminated using the Grassmannian Novikov theorem (see Ref.~\cite{zhao_fermionic_2012} and~\ref{sec:derivation_master}).
Finally, we obtain the following  hierarchy of density operators
\begin{eqnarray}
  \partial_t \rhot[\mm,\nn]
  &=& -\ii \Comm{H, \rhot[\mm,\nn]} - \left( \mm\cdot\ww + \nn\cdot\cc\ww \right) \rhot[\mm,\nn] \nonumber\\
  &&+ \sum_j (-1)^{\abs{\mm}_j} g_j L_j \rhot[\mm-\ee_j,\nn] + \sum_j (-1)^{\abs{\nn}_j} \cc{g}_j \rhot[\mm,\nn-\ee_j] \adj{L}_j \nonumber\\
  &&- \sum_j \left( (-1)^{\abs{\mm}_j} \adj{L}_j \rhot[\mm+\ee_j,\nn] - (-1)^{\abs{\nn}} \rhot[\mm+\ee_j,\nn] \adj{L}_j \right) \nonumber\\
  &&+ \sum_j \left( (-1)^{\abs{\mm}} L_j \rhot[\mm,\nn+\ee_j] - (-1)^{\abs{\nn}_j} \rhot[\mm,\nn+\ee_j] L_j \right)
\label{eq:master_hierarchy}
\end{eqnarray}

This constitutes the second result of the present work.
As for the stochastic fermionic hierarchy \eqref{eq:pure_hierarchy}, this is a finite set of $2^{2J}$ equations.
For numerical purposes it might again be advantageous to use a suitable truncation.
Also, the accuracy of a truncated hierarchy can be increased by approximating truncated auxiliary operators using a so called terminator~\cite{sus_hierarchy_2014}.

\subsection{Thermal initial state}
\label{sub:thermal}

Clearly, the pure state hierarchy~\eqref{eq:pure_hierarchy}---and therefore also the density operator hierarchy~\eqref{eq:master_hierarchy}---depends crucially on the bounded domains of the memory integrals.
These only arise for an initial vacuum state of the bath~\eqref{eq:initial_Psi} as indicated in the footnote on page~\pageref{fn:integral_boundaries}.
Remarkably, it is possible to map the equations of motion corresponding to an initial thermal state
\[
  \rho_{\rm tot}(t=0) = \ket{\psi_0}\bra{\psi_0} \otimes \rho_{\rm th}
\]
to the established zero-temperature NMQSD equation~\eqref{eq:nmsse}.
Here, the thermal bath state is given by
\(
  \rho_{\rm th}=\exp[-\frac{H_{\rm env}-\mu N_{\rm env}}{T}]/\mathcal{Z},
\)
with the chemical potential $\mu$ and the partition function
\(
  \mathcal{Z}=\Tr_{\rm env} \, \exp[-\frac{H_{\rm env}-\mu N_{\rm env}}{T}].
\)
By doubling the degrees of freedom using the well-known Bogoliubov transformation~\cite{diosi_non-markovian_1998,yu_non-markovian_2004,ritschel_non-markovian_2014} the resulting NMQSD equation reads~\cite{chen_non-markovian_2013}
\begin{eqnarray}
  \label{eq:nmsse_thermal}
  \partial_t \psit
  &=& -\ii H \psit
  + \sum_j L_j \ZZ_j(t) \psit + \sum_j \adj{L}_j \cc{W}_j(t) \psit \nonumber\\
  &&- \sum_j \adj{L}_j \int_0^t \alpha_j(t-s) \frac{\delta \psit}{\delta\ZZ_j(s)} \dd s\\
  &&- \sum_j L_j \int_0^t \beta_j(t-s) \frac{\delta \psit}{\delta\cc{W}_j(s)} \dd s \nonumber,
\end{eqnarray}
where we now have another auxiliary process $\cc{W}_j(t)$ for each original process $\cc{Z}_j(t)$.
The correlation functions $\alpha_j(t)$ and $\beta_j(t)$ in~\eqref{eq:nmsse_thermal} also characterize these noise processes:
\begin{eqnarray}
  \E[ Z_j(t) \ZZ_j(s) ] = \alpha_j(t-s) = \int_0^\infty J_j(\omega) (1 - \bar{n}_j(\omega)) \exp[-\ii\omega (t-s)] \,\dd\omega \\
  \E[ W_j(t) \cc{W}_j(s) ] = \beta_j(t-s) = \int_0^\infty J_j(\omega) \bar{n}_j(\omega) \exp[\ii\omega (t-s)] \,\dd\omega,
  \label{eq:sum_process_correlations}
\end{eqnarray}
with all other relations being zero similar to~\eqref{eq:correlations}.
Here, $\bar{n}_j(\omega)$ denotes the Fermi-Dirac distribution function (now with a possibly different chemical potential for each independent bath)
\(
  \bar{n}_j(\omega) = (\exp[(\omega - \mu_j) / T] + 1)^{-1}.
\)

In the case of self-adjoint coupling operators $L_j = \adj{L}_j$ one can go on step further and combine each $\ZZ_j(t)$ and $\cc{W}_j(t)$ into the sum processes
\(
  \tilde\ZZ_j(t) = \ZZ_j(t) + \cc{W}_j(t).
\)
Remarkably, the corresponding correlation function
\[
  \tilde\alpha(t) = \alpha(t) + \beta(t)
  = \int_0^{\infty} J(\omega)\Big\{\cos(\omega t) - \ii \tanh\big(\frac{\omega}{2 T}\big)\sin(\omega t) \Big\} \dd\omega
  \label{eq:bcf}
\]
is also the well known thermal correlation function for a spin bath~\cite{weiss_quantum_2012}.
In conclusion, the resulting finite temperature NMQSD equation in this case is identical to the zero-temperature version~\eqref{eq:nmsse} except for the thermal bath correlation function.\\

Besides incorporating the effects of finite temperature on the system's dynamics, the thermal bath correlation function~\eqref{eq:bcf} also provides a natural way to obtain the crucial decomposition of the BCF as a sum of exponentials.
Here, we will only sketch the idea following the detailed exposition for a bosonic environment in Ref.~\cite{RiEi14_094101_}:
Due to the symmetric behavior under reflection at the origin of the term in braces in~\eqref{eq:bcf}, a symmetric continuation
\[
  \tilde J (\omega) = \left\{
    \begin{array}{lr}
      J(\omega) &: \omega \ge 0 \\
      J(-\omega) &: \omega < 0
    \end{array}
  \right.
\]
lets us expand the integral boundaries in~\eqref{eq:bcf} to the whole real axis
\[
  \alpha(t) = \frac{1}{2} \int_{-\infty}^{\infty} \tilde J(\omega) \left\{ \cos(\omega t) - \ii \tanh( \frac{\omega}{2 T} ) \sin(\omega t)  \right\} \,\dd \omega.
\]
Closing the integral contour in the upper/lower complex half plane and employing the residual theorem yields exactly the sought after sum of exponentials provided we can express the integrand as a sum of poles.
Many realistic spectral are given as a finite sum of poles~\cite{ritschel_non-markovian_2014}, whereas for the hyperbolic tangents a suitable sum-over-poles scheme has to be employed, i.e.\ the Matsubara decomposition~\cite{mahan_many-particle_2000}, continued fraction expansion~\cite{ozaki_continued_2007}, or the Padé decomposition~\cite{hu_communication:_2010}.

\section{Bosonic Environments}
\label{sec:bosonic_environments}

The second part of the paper is devoted to the related master equation hierarchy for an environment of harmonic oscillators.
The bosonic microscopical Hamiltonian is identical to~\eqref{eq:h_tot}---the Hamiltonian used in the previous section---except that the environment's creation and annihilation operators $\adj{b}_{j,\lambda}$ and $b_{j,\lambda}$ are replaced by their bosonic counterparts.
In other words, the anti-commutators in Eq.~\eqref{eq:anticommutators} are replaced by commutators
\[
  \big[ b_{j,\lambda}, b_{j',\lambda'} \big] = 0
  \quad\mbox{and}\quad
  \big[ b_{j,\lambda}, \adj{b}_{j',\lambda'} \big] = \delta_{jj'} \delta_{\lambda\lambda'}.
\]
Without any approximation, such a model leads to the well-known NMQSD equation~\cite{strunz_linear_1996,Di96_309_,diosi_non-markovian_1997,diosi_non-markovian_1998}, which agrees with Eq.~\eqref{eq:nmsse} except for the noise $\ZZ$:
In contrast to the fermionic case, the noise process in the bosonic equation is complex valued and, therefore, efficiently implementable for large classes of bath correlation functions~\cite{garcia-ojalvo_noise_1999}.

\subsection{Bosonic Hierarchy of Pure States}
\label{sub:bosonic_hierarchy_of_pure_states}

Due to the similarities of the underlying NMQSD equations, the fermionic (Eq.~\eqref{eq:pure_hierarchy}) and bosonic hierarchy of pure states~\cite[Eq.~(14)]{sus_hierarchy_2014}
\begin{eqnarray}
  \partial_t \psit[\kk]
  &=& \left( -\ii H - \kk\cdot\ww + \sum_j \ZZ_j(t) L_j \ \right) \psit[\kk] \nonumber\\
  &&+ \sum_j  g_j L_j \psit[\kk - \ee_j]
  - \sum_j  \adj{L}_j \psit[\kk + \ee_j]
  \label{eq:hops}
\end{eqnarray}
are remarkably similar.
Besides the obvious sign prefactors in~\eqref{eq:pure_hierarchy} and the different noise processes, the crucial difference is number of auxiliary states:
Whereas the fermionic hierarchy is always finite due to the condition $k_j \in \{0,1\}$, its bosonic counterpart is originally infinite with $k_j \in \mathbb{N}_0$.
Only a suitable truncation turns the latter into a practical scheme~\cite{sus_hierarchy_2014}.

\subsection{Bosonic Master Equation Hierarchy}
\label{sub:bosonic_master_equation_hierarchy}

Clearly, the similarities of the two pure state hierarchies carries over to the hierarchy of density matrices:
The construction of the \emph{bosonic master equation hierarchy}
\begin{eqnarray}
  \label{eq:master_bosonic}
  \partial_t \rhot[\mm,\nn]
  &=& -\ii \Comm{H, \rhot[\mm,\nn]} - \left( \mm\cdot\ww + \nn\cdot\cc\ww \right) \rhot[\mm,\nn] \nonumber\\
  &&+ \sum_j m_j g_j L_j \rhot[\mm-\ee_j,\nn] + \sum_j n_j \cc{g}_j \rhot[\mm,\nn-\ee_j] \adj{L}_j \\
  &&- \sum_j \left[ \adj{L}_j, \rhot[\mm+\ee_j,\nn] \right] + \sum_j \left[ L_j, \rhot[\mm,\nn+\ee_j] \right] \nonumber
\end{eqnarray}
runs along the same lines as the derivation in Sec.~\ref{sub:fermionic_hierarchy}.
Here, the indices $m_j, n_j \in \mathbb{N}_0$ are not bounded from above and, therefore, the hierarchy is infinite in principle and has to be truncated for numerical simulations.
Note the close relation between the fermionic and bosonic equations: besides the prefactors $m_j$ and $n_j$ in~\eqref{eq:master_bosonic}, which are irrelevant for $m_j, n_j \in \{0,1\}$, the two hierarchies only differ in the additional sign factors $(-1)^\abs{\mm}$ and $(-1)^\abs{\nn}$ in~\eqref{eq:master_hierarchy}.

\section{Conclusions}
\label{sec:Conclusions}
In the present work we have considered an open system model with a fermionic environment which is coupled linearly to the system.
The two main results are the derivation of a hierarchy of pure states~\eqref{eq:pure_hierarchy} and the corresponding hierarchy for density matrices~\eqref{eq:master_hierarchy}.
The starting point was the general stochastic NMQSD equation that contains Grassmannian noise as well as a functional derivatives with respect to this noise under a memory integral.
The hierarchy of pure states~\eqref{eq:pure_hierarchy} no longer contains this functional derivative, however, it still contains the Grassmann noise, which hinders efficient numerical simulations.
In contrast, the  hierarchy for density matrices~\eqref{eq:master_hierarchy} also gets rid of the noise making it suitable for numerical simulations.

Both, the hierarchy of pure states \eqref{eq:pure_hierarchy} and the  hierarchy for density matrices \eqref{eq:master_hierarchy}, are actually finite and allow to compute the reduced density operator of the system exactly.
However, the number of coupled differential equations scales as $2^{J}$ for the pure state hierarchy and as  $2^{2 J}$ for the hierarchy for density matrices.
Note that the objects entering these equations have the dimension $D$ of the system Hilbert space in the case of the pure state hierarchy and $D^2$ for the matrix hierarchy.
Since for a large number of modes $J$ the size of the problem becomes numerically intractable, it is necessary to truncate the hierarchy in a suitable way.
This can be done along the lines discussed at the end of Sec.~\ref{sub:fermionic_hierarchy}.

In the second part of the paper we showed that the construction of a density operator hierarchy from a hierarchy of pure states is also feasible for bosonic environments.
Since in this case, both, Eq.~\eqref{eq:hops} and~\eqref{eq:master_bosonic} are suitable for numerical simulations, a comparison of their performance is needed to asses their respective strengths and weaknesses [to be published].
Furthermore, there remains the challenging problem to establish the connection between the density operator hierarchies based on HOPS presented here and the established hierarchical equations of motion~\cite{JiZhYa08_234703_,tanimura_stochastic_2006}.

It is even possible to treat systems coupled to both types of environments using the approach presented in this paper.
Also, further generalizations to multiple distinguishable kinds of fermions or to a spin bath (e.g.\ arising as low-temperature limit of localized modes~\cite{prokofev_theory_2000}) are possible as well.
This makes the hierarchical approach presented in this a paper a highly flexible tool in the field of open quantum system dynamics.

\ack
DS acknowledges support by the Excellence Initiative of the German Federal and State Governments (Grant ZUK 43), the ARO under contracts W911NF-14- 1-0098 and W911NF-14-1-0133 (Quantum Characterization, Verification, and Validation), and the DFG.

\appendix

\section{Derivation of the fermionic hierarchy of pure states}
\label{sec:derivation_of_hierarchy}
The derivation for the fermionic hierarchy of pure states is very similar to that of the bosonic one discussed in \cite{sus_hierarchy_2014}
To have a compact notation, we ignore the condition $k_j \in \{0,1\}$ for our derivation.
In the end it will turn out that these conditions are trivially incorporated due to the structure of the hierarchy.

We start by taking the time derivative of $\psit[\kk]$.
Using its definition \eqref{eq:aux_states} we find
\begin{eqnarray}
  \partial_t \psit[\kk]
  &=& ( \partial_t \DDeriv{t}^{\kk} ) \psit + \DDeriv{t}^{\kk} (\partial_t \psit)
\end{eqnarray}
For the first term on the right hand side we use\footnote{%
  Note that the bounded integral domain in the memory integral that appears in the final equation is due to vacuum initial conditions~\eqref{eq:initial_Psi}.
}
\(
  (\partial_t\Deriv[j]{t}) \psit = - w_j \Deriv[j]{t} \psit
\).
For the second term on the right hand side we use that  all system operators commute with all $\Deriv[j]{t}$ and obtain
\begin{eqnarray}
\partial_t \psit[\kk]
  &=& - \kk\cdot\ww \, \psit[\kk] \nonumber\\
  && - \ii H \psit[\kk]+ \underbrace{\sum_j L_j \DDeriv{t}^{\kk} \ZZ_j(t) \psit}_{(*)}
  - \underbrace{\sum_j \adj{L}_j \DDeriv{t}^{\kk} \Deriv[j]{t} \psit}_{(**)}
\end{eqnarray}
To obtain a closed equation for the auxiliary states,  we want the $\Deriv[j]{t}$ ordered as in the definition \eqref{eq:aux_states}.
In $(**)$ we have to move $\Deriv[j]{t}$ to the correct position  (note the ordering in~\eqref{eq:aux_states}):
\begin{eqnarray}
  \DDeriv{t}^{\kk} \Deriv[j]{t}
  &=& (-1)^{k_J} \Deriv[1]{t}^{k_1} \ldots \Deriv[j]{t} \Deriv[J]{t}^{k_J} \nonumber\\
  &=& (-1)^{k_{j+1} + \cdots + k_J} \Deriv[1]{t}^{k_1} \ldots \Deriv[j]{t}^{k_j + 1} \cdots \Deriv[J]{t}^{k_J}.
  \label{eq:derivation_deriv_term}
\end{eqnarray}
In $(*)$ we have to bring  $\ZZ_j(t)$ in front of $\DDeriv{t}^{\kk}$.
This can be achieved by noting that
\(
  \{ \Deriv[j]{t}, \ZZ_{j'}(s) \} = \delta_{jj'} \, \alpha(t-s)
\).
We then find
\begin{eqnarray}
  \DDeriv{t}^{\kk} \ZZ_j(t)
  &=& (-1)^{k_J} \Deriv[1]{t}^{k_1} \ldots \ZZ_j(t) \Deriv[J]{t}^{k_J} \nonumber\\
  &=& (-1)^{k_{j+1} + \cdots + k_J} \Deriv[1]{t} \ldots \Deriv[j]{t}^{k_j} \ZZ_j(t) \ldots \nonumber\\
  &=& (-1)^{\abs{\kk}_j} \Deriv[1]{t} \ldots  \left( -\Deriv[j]{t}^{k_j-1} \ZZ_j(t) \Deriv[j]{t} + \Deriv[j]{t}^{k_j-1} g_j \right) \ldots \nonumber\\
  &=& \ldots \left( \Deriv[j]{t}^{k_j-2} \ZZ_j(t) \Deriv[j]{t}^2 - \Deriv[j]{t}^{k_j-1} g_j + \Deriv[j]{t}^{k_j-1} g_j \right) \ldots \nonumber\\
  &=& \ldots \left( (-1)^{k_j} \ZZ_j(t) \Deriv[j]{t}^{k_j} + (k_j \bmod 2) g_j \Deriv[j]{t}^{k_j-1} \right) \ldots \nonumber\\
  &=& (-1)^\abs{\kk} \ZZ_j(t) \DDeriv{t}^{\kk} + (-1)^{\abs{\kk}_j} (k_j \bmod 2) g_j \DDeriv{t}^{\kk - \ee_j}
  \label{eq:derivation_noise_term}
\end{eqnarray}
where $\abs{\kk}$ and $\abs{\kk}_j$ have been defined below Eq.~\eqref{eq:pure_hierarchy}.

Combining~\eqref{eq:derivation_deriv_term} and~\eqref{eq:derivation_noise_term} leads to the (apparently) infinite hierarchy of pure states for fermionic environment
\begin{eqnarray}
  \partial_t \psit[\kk]
  &=& \left( -\ii H - \kk\cdot\ww + (-1)^\abs{\kk} \sum_j \ZZ_j(t) L_j \right) \psit[\kk] \nonumber\\
  &&+ \sum_j (-1)^{\abs{\kk}_j} (k_j \bmod 2) g_j L_j \psit[\kk - \ee_j] \nonumber\\
  &&- \sum_j (-1)^{\abs{\kk}_j} (-1)^{\abs{\kk}_j} \adj{L}_j \psit[\kk + \ee_j]
  \label{eq:pure_hierarchy_tmp}
\end{eqnarray}
Note that all states with some $k_j \notin \{0,1\}$---which should be zero actually---only couple to other states also satisfying this condition:
Due to the modulo function in the term coupling to states \quotes{below} in the hierarchy, states with some $k_j \notin \{0,1\}$ that are initially zero always remain zero.
Therefore, the closed and finite hierarchy with all $k_j \in \{0,1\}$ and equation \eqref{eq:pure_hierarchy_tmp} can be written as~\eqref{eq:pure_hierarchy}.

\section{Derivation of Master Equation hierarchy}
\label{sec:derivation_master}
In this appendix we provide the Novikov theorem, which is essential to get from Eq.~\eqref{eq:master_hierarchy_almost} to Eq.~\eqref{eq:master_hierarchy}.
The Novikov theorem allows us to get rid of the explicit dependence of the Grassmann processes in the second and third line of~\eqref{eq:master_hierarchy_almost} by a \quotes{partial integration}.
For the fermionic case the Novikov theorem has been discussed in \cite{zhao_fermionic_2012} (see Eq.~(22) and (23) therein).
We need two variants of the Novikov theorem:
\begin{eqnarray}
  \E[ \ket{\psit[\mm]}\bra{\npsit[\nn]} Z_j(t) ]
  &=& - \E \Bigg( \int \mathrm{d}s \, \alpha_j(t-s) \frac{\overrightarrow\delta}{\delta \ZZ_j(s)} \ket{\psit[\mm]}\bra{\npsit[\nn]} \Bigg) \nonumber\\
  &=& - \rhot[\mm+\ee_j,\nn]
  \label{eq:Novokov_right}
\end{eqnarray}
and
\begin{eqnarray}
  \E[ \ZZ_j(t) \ket{\psit[\mm]}\bra{\npsit[\nn]} ]
  &=& - \E \Bigg( \int \mathrm{d}s \, \cc{\alpha_j(t-s)} \ket{\psit[\mm]}\bra{\npsit[\nn]} \frac{\overleftarrow\delta}{\delta Z_j(s)} \Bigg) \nonumber\\
  &=& \rhot[\mm,\nn+\ee_j],
  \label{eq:Novokov_left}
\end{eqnarray}
where in the second line of each equation we have used the definition of the auxiliary matrices \eqref{eq:def_rho} and the definitions \eqref{eq:derivation_operator} and \eqref{eq:aux_states}.
In the second equation the right-functional derivative appears and we have used
\[
  \frac{\overrightarrow\delta \npsit(\ZZ)}{\delta \ZZ_j(s)} = \frac{\overrightarrow\delta \psit(-\ZZ)}{\delta\ZZ_j(s)} = - \left. \frac{\overrightarrow\delta \psit(\cc{Z'})}{\delta\cc{Z_j'}(s)} \right|_{\cc{Z'} = -\ZZ}.
\]
These two equations~\eqref{eq:Novokov_right} and~\eqref{eq:Novokov_left} show that it is possible to express the averages in the second and third line of~\eqref{eq:master_hierarchy_almost} containing the noise process explicitly by the auxiliary density operators.

\vspace{1cm}

\bibliographystyle{journal_v5}
\bibliography{references_ae,references_ds}
\end{document}